**Glassy behavior in the ferromagnetic and the non-magnetic insulating states of the rare earth manganates, $Ln_{0.7}Ba_{0.3}MnO_3$ (Ln = Nd or Gd)**

Asish K. Kundu[1,2], P. Nordblad[2] and C. N. R. Rao[1]

[1]Chemistry and Physics of Materials Unit, Jawaharlal Nehru Centre for Advanced Scientific Research, Jakkur P.O., Bangalore-562064, India,

[2]Department of Engineering Sciences, Uppsala University, 751 21 Uppsala, Sweden

## Abstract

While $La_{0.7}Ba_{0.3}MnO_3$ is a ferromagnetic metal ($T_C$ = 340 K) with long-range ordering, $Nd_{0.7}Ba_{0.3}MnO_3$ shows a transition around 150 K with a small increase in magnetization, but remains an insulator at all temperatures. $Gd_{0.7}Ba_{0.3}MnO_3$ is non-magnetic and insulating at all temperatures. Low-field dc magnetization and ac susceptibility measurements reveal the presence of the transition around 150 K in $Nd_{0.7}Ba_{0.3}MnO_3$, and a complex behavior with different ordering/freezing transitions at 62, 46 and 36 K in the case of $Gd_{0.7}Ba_{0.3}MnO_3$, the last one being more prominent. The nature of the field-dependence of the magnetization combined with the slow magnetic relaxation, aging and memory effects suggest that $Nd_{0.7}Ba_{0.3}MnO_3$ is a cluster-glass below 150 K, a situation similar to that found in $La_{1-x}Sr_xCoO_3$. $Gd_{0.7}Ba_{0.3}MnO_3$, however, shows non-equilibrium dynamics characteristic of spin-glasses, below 36 K. The difference in the nature of the glassy behavior between $Gd_{0.7}Ba_{0.3}MnO_3$ and $Nd_{0.7}Ba_{0.3}MnO_3$ probably arises because of the larger disorder arising from the mismatch between the sizes of the A-site cations in the former. Our results on $Nd_{0.7}Ba_{0.3}MnO_3$ and $Gd_{0.7}Ba_{0.3}MnO_3$ suggest that the magnetic insulating states often reported in the rare earth manganates of the type $Ln_{1-x}A_xMnO_3$ (Ln = rare earth, A = alkaline earth) are likely to be associated with glassy magnetic behavior.

## 1. INTRODUCTION

Among the several novel properties and phenomena exhibited by the rare earth manganates of the type $Ln_{1-x}A_xMnO_3$ (Ln = rare earth, A = alkaline earth), charge ordering and electronic phase separation are of particular interest [1-6]. Both these properties are highly sensitive to the average radius of the A-site cations, $\langle r_A \rangle$, and the size-disorder arising from the mismatch between the A-site cations [4-7]. The size-disorder is generally expressed in terms of the $\sigma^2$ parameter which is defined as, $\sigma^2 = \sum x_i r_i^2 - \langle r_A \rangle^2$, where $x_i$ is the fractional occupancy of A-site ions, $r_i$ is the corresponding ionic radii and $\langle r_A \rangle$ is the weighted average radius calculated from $r_i$ values [7]. Electronic phase separation is found to occur above a critical composition $x_c$ in $La_{0.7-x}Ln_xCa_{0.3}MnO_3$, specially in the regime when $\langle r_A \rangle$ is close to 1.18 Å or lower, and is favored by large size-disorder [3, 6]. In this system, $\langle r_A \rangle$ decreases with increasing x, affecting the $e_g$ bandwidth. A study of $La_{0.250}Pr_{0.375}Ca_{0.375}MnO_3$ by Deac et al. [8] has shown two types of magnetic relaxation, one at low fields associated with the reorientation of ferromagnetic (FM) domains and another at higher fields due to the transformation between FM and non-FM phases. The presences of FM-clusters and associated magnetic relaxation phenomena well below $T_C$ have been reported in $La_{0.7-x}Y_xCa_{0.3}MnO_3$ by Freitas et al. [9]. $Nd_{0.7}Sr_{0.3}MnO_3$ with a well defined ferromagnetic $T_C$ exhibits aging phenomena in the ferromagnetic phase indicating magnetic frustration and disorder [10]. Lopez et al. [11]

have provided evidence for two competing magnetic phases in $La_{0.5}Ca_{0.5}MnO_3$ based on a magnetic relaxation study. A recent investigation of the magnetic and electric properties of $La_{0.7-x}Ln_xBa_{0.3}MnO_3$ (Ln = Pr, Nd, or Gd) where the $\langle r_A \rangle$ remains relatively large over the entire range of compositions ($\geq$ 1.216 Å) has shown that the FM or non-magnetic insulating compositions can be rendered FM and metallic by decreasing the size disorder [12]. An insulating magnetic state is found in $Nd_{0.7}Ba_{0.3}MnO_3$, but $Gd_{0.7}Ba_{0.3}MnO_3$ is insulating and non-magnetic down to low temperatures, although the carrier concentration ($Mn^{3+}/Mn^{4+}$ ratio) is the same as in $La_{0.7}Ba_{0.3}MnO_3$ which is a genuine FM metal [12, 13]. In this paper, we focus our interest on $Nd_{0.7}Ba_{0.3}MnO_3$ and $Gd_{0.7}Ba_{0.3}MnO_3$ and have carried out a detailed study of the magnetic properties by employing measurements of low-field dc magnetization, ac susceptibility, magnetic relaxation and memory effects. The study has revealed the presence of glassy magnetic phases in both these manganates, albeit of different varieties.

## 2. EXPERIMENTAL PROCEDURE

Polycrystalline samples of $Ln_{0.7}Ba_{0.3}MnO_3$ (Ln = Nd and Gd) were prepared by the conventional solid-state reactions. Stoichiometric mixtures of the respective rare earth oxides, $BaCO_3$ and $MnO_2$ were weighed in the desired proportions and milled for few hours with propanol. The mixtures were dried, and calcined in air at 1223 K followed by heating at 1273 K and 1373 K for 12h each in air. The powders thus obtained were pelletized and the pellets sintered at 1673 K for 24h in air. Composition analysis was carried out using Energy Dispersive X-ray (EDX) analysis using a LEICA S440I scanning electron microscope fitted with a Si-Li detector and it confirms the composition within experimental errors. The oxygen stoichiometry was determined by iodometric titrations.

The phase purity of the manganates was established by recording the X-ray diffraction patterns in the 2θ range of 10°-80° with a Seiferts 3000 TT diffractometer using Cu-Kα radiation. A Quantum Design MPMSXL superconducting quantum interference device (SQUID) magnetometer and a non-commercial low field SQUID magnetometer system [14] were used to investigate the magnetic properties of the samples. The temperature dependence of the zero-field-cooled (ZFC) and field-cooled (FC) magnetization was measured in different applied magnetic fields. Hysteresis loops were recorded at some different temperatures in the low temperature phases of the system. The dynamics of the magnetic response was studied by ac-susceptibility measurements at different frequencies and measurements of the relaxation of the low field ZFC magnetization.

In the measurements of the temperature dependence of the ZFC magnetization, the sample was cooled from 350 K to 10 K in zero-field, the field was applied at 10 K and the magnetization recorded on re-heating the sample. In the FC measurements the sample was cooled in the applied field to 10 K and the magnetization recorded on re-heating the sample, keeping the field applied. In the relaxation experiments, the sample was cooled in zero-field from a reference temperature of 170 K (for Nd) and 90 K (for Gd) to a measuring temperature, $T_m$ and kept there during a wait time, $t_w$. After the wait time, a small probing field was applied and the magnetization was recorded as a function of time elapsed after the field application. The electrical resistivity (ρ) measurements were carried out by a standard four-probe method with silver epoxy as electrodes in the 20-300 K temperature range.

## 3. RESULTS AND DISCUSSION

$La_{0.7}Ba_{0.3}MnO_3$, $Nd_{0.7}Ba_{0.3}MnO_3$ and $Gd_{0.7}Ba_{0.3}MnO_3$ possess orthorhombic structures (*Pnma* space group) and the lattice parameters decrease with the decrease in the size of the rare earth ion as expected. In figure 1, we show the variation of lattice parameters and cell volume with $<r_A>$ to demonstrate this feature. The $<r_A>$ values of $La_{0.7}Ba_{0.3}MnO_3$, $Nd_{0.7}Ba_{0.3}MnO_3$ and $Gd_{0.7}Ba_{0.3}MnO_3$ are 1.292, 1.255 and 1.216 Å respectively, the corresponding values of the size-disorder parameter, $\sigma^2$, being 0.014, 0.020 and 0.027 Å$^2$ respectively. Thus, $Gd_{0.7}Ba_{0.3}MnO_3$ has the smallest $<r_A>$ and the largest $\sigma^2$.

In figure 2(a), we show the dc magnetization behavior of $La_{0.7}Ba_{0.3}MnO_3$, $Nd_{0.7}Ba_{0.3}MnO_3$ and $Gd_{0.7}Ba_{0.3}MnO_3$ under FC conditions (H = 500 Oe). $La_{0.7}Ba_{0.3}MnO_3$ shows a sharp increase in the magnetization around 340 K ($T_C$) corresponding to the ferromagnetic transition. There is evidence for saturation, the values of the saturation magnetization and the corresponding magnetic moment being 35 emu/g and 1.5 $\mu_B$/f.u. $Nd_{0.7}Ba_{0.3}MnO_3$ shows an increase in the magnetization around 150 K, but the maximum magnetization value found is 18 emu/g (0.8 $\mu_B$/f.u.) at 40 K. $Gd_{0.7}Ba_{0.3}MnO_3$ shows no evidence for a magnetic transition and the magnetization value is 5 emu/g (0.25 $\mu_B$/f.u.) at 40 K. Clearly, the magnetic properties of the three manganates are distinctly different from one another. Whereas $La_{0.7}Ba_{0.3}MnO_3$ shows metallic behavior below $T_C$ , $Nd_{0.7}Ba_{0.3}MnO_3$ and $Gd_{0.7}Ba_{0.3}MnO_3$ show insulating behavior over the entire temperature range (figure 2(b)). Thus, $Nd_{0.7}Ba_{0.3}MnO_3$ is insulating at and below the 150 K transition and $Gd_{0.7}Ba_{0.3}MnO_3$ is a non-magnetic insulator at all temperatures.

Magnetization data of $La_{0.7}Ba_{0.3}MnO_3$ at low fields were similar to those obtained at higher field showing little divergence between the ZFC and FC data. In figure 3, we present the low-field ZFC and FC magnetization data of $Nd_{0.7}Ba_{0.3}MnO_3$ and $Gd_{0.7}Ba_{0.3}MnO_3$. The FC magnetization of $Nd_{0.7}Ba_{0.3}MnO_3$ shows the transition around 150 K. $Gd_{0.7}Ba_{0.3}MnO_3$ exhibits a rather complex behavior below 62 K where irreversibility between the ZFC and FC magnetization data first appears (figure 3(c)). The low temperature region is discussed later, but it is noteworthy that there are three characteristic temperatures: 62 K (onset of significant irreversibility between the ZFC and FC magnetization curves), 46 K (a maximum in the FC curve) and 36 K (a maximum in the ZFC curve), all indicating different ordering and/or freezing processes in the system.

Figure 4 shows the field-variation of magnetization at three different temperatures for $Nd_{0.7}Ba_{0.3}MnO_3$ and $Gd_{0.7}Ba_{0.3}MnO_3$. Both these manganates do not exhibit hysteresis. Below 150 K, $Nd_{0.7}Ba_{0.3}MnO_3$ shows a behavior similar to a weak ferromagnet, the magnetization approaching saturation at high fields. $Gd_{0.7}Ba_{0.3}MnO_3$ does not show the M-H behavior of a ferromagnet at low temperatures, and exhibits no tendency for saturation even at high fields. The shape of the M-H curve and the absence of saturation even at high fields in $Gd_{0.7}Ba_{0.3}MnO_3$ are reminiscent of magnetization curves of spin glasses [15]. The M-H behavior becomes nearly linear (paramagnetic) at 200 K in both the manganates.

The temperature dependence of the ac susceptibility of $Nd_{0.7}Ba_{0.3}MnO_3$ and $Gd_{0.7}Ba_{0.3}MnO_3$ is presented at different frequencies in figure 5. The in-phase $\chi'(T)$ component of the ac susceptibility reveals similar features as the ZFC-magnetization at low-fields in both the manganates. $Nd_{0.7}Ba_{0.3}MnO_3$ shows a sharp maximum below 150 K, which is frequency-independent. However, there is a weak frequency dependence at temperatures below 140 K, a behavior noted earlier in $Nd_{0.7}Sr_{0.3}MnO_3$ [10]. $Gd_{0.7}Ba_{0.3}MnO_3$ shows a shoulder around 62 K, a weak anomaly just above 46 K and a maximum at 36 K. The $\chi'(T)$ data become strongly frequency-dependent below 36 K. This transition could arise from the presence of small magnetic clusters in a non-magnetic matrix just as in cobaltates of the type $La_{0.7}Ca_{0.3}CoO_3$ [16]. Other examples of oxide systems where only such short-range ferromagnetic correlations occur are known [10, 16, 17].

Time-dependent ZFC magnetization measurements show that both $Nd_{0.7}Ba_{0.3}MnO_3$ and $Gd_{0.7}Ba_{0.3}MnO_3$ exhibit logarithmically slow dynamics and aging at low temperatures. In figures 6(a) and (b), we show the time-dependent ZFC magnetization, m(t), measured at $T_m = 40$ K, and the corresponding relaxation rates $S(t) = 1/H\ [dM_{ZFC}(T,t_w,t)/d\log_{10}(t)]$ for $Nd_{0.7}Ba_{0.3}MnO_3$. The applied field was 1 Oe and the wait times were $t_w = 100$, 1000 and 10000 s. The results of similar measurements on $Gd_{0.7}Ba_{0.3}MnO_3$ at 30 K are presented in figures 7(a) and (b). The wait-time dependence of the magnetic relaxation illustrated in figures 6 and 7 shows that both the manganates are subject to magnetic aging at low temperatures. Relaxation experiments (not shown) at 80 K (Nd) and 40 K (Gd) reveal slow relaxation and aging behavior at these temperatures as well, but with a much decreased relaxation rate compared to that at low temperatures.

Time-dependent thermo-remanent magnetization (TRM) measurements at the same temperatures yielded similar results for both systems. Magnetic aging is a signature of spin-glasses [15] and explained within the droplet (or domain growth) model the maximum in the relaxation rate is associated with a crossover between quasi-equilibrium and non-equilibrium dynamics [18]. The slow relaxation and aging behavior of $Nd_{0.7}Ba_{0.3}MnO_3$ and $Gd_{0.7}Ba_{0.3}MnO_3$ demonstrate that magnetic disorder and frustration occur in the low-temperature phases.

Glassy dynamics in spin glasses is also manifested by a memory effect that can be demonstrated by dc-magnetization or low frequency ac-susceptibility experiments. We have employed zero-field-cooled magnetization vs. temperature experiments [16] to investigate possible memory phenomena in the two manganates. The experiment includes a reference measurement, according to the ZFC protocol described earlier, and a similar ZFC memory experiment, the protocol of which only includes one additional feature, the cooling down of the sample is halted at a stop temperature for some hours. In a spin glass experiment, the memory curve acquires a weak dip at the temperature where the zero-field cooling was halted. To illustrate the memory effect, it is convenient to plot the difference between the reference and the memory curve. A spin glass phase (ordinary or re-entrant) has a pronounced memory behavior, whereas a disordered and frustrated ferromagnetic phase shows little or no memory effect. In the case of $Nd_{0.7}Ba_{0.3}MnO_3$, we carried out the ZFC experiment by cooling the sample from a reference temperature of 170 K to 90 K, where the magnetic field (10 Oe) was applied and the magnetization recorded on continuously heating the sample to 170 K. The ZFC memory curve was recorded in a similar way with the additional feature that the cooling in zero field was stopped at 120 K for 3 hours. Figure 8(a) shows the two curves. A weak dip can barely be discerned in the memory curve (labelled 120 K). The difference plot $M_{mem}$-$M_{ref}$ versus T, shown as an inset in figure 8(a) reveals a broad but shallow memory of the stop at 120 K. In contrast, the corresponding experiment on the $Gd_{0.7}Ba_{0.3}MnO_3$ sample shows a prominent memory dip. The experiment was performed starting from 70 K and cooling the sample continuously to 20 K, with an intermediate stop at 30 K for 3 hours in the memory measurement. Figure 8(b) shows the two curves. There is a significant difference between the reference and the memory curves. The difference plot shown in the inset of figure 8(b) reveals a deep, broad memory dip. The dip abruptly ceases above 36 K. The memory behavior of $Gd_{0.7}Ba_{0.3}MnO_3$ at 36 K is clearly that of a spin-glass.

## 4. CONCLUSIONS

$Nd_{0.7}Ba_{0.3}MnO_3$ shows an increase in magnetization at 150 K, but the value of magnetization is small at low temperature. It is also an insulator. It shows a pronounced aging behavior, but a rather weak memory effect below 150 K, probably due to the presence of FM clusters in an insulating matrix. $Nd_{0.7}Ba_{0.3}MnO_3$ appears to be a cluster glass or a magnetically disordered system similar to $La_{1-x}Sr_xCoO_3$ [19]. $Gd_{0.7}Ba_{0.3}MnO_3$ appears to contain small magnetic clusters, giving rise to a spin-glass state below 36 K. Low-field magnetization experiments indicate that some kind of ordering/or freezing process occurs in this manganate even around 62 K, with an additional process at 46 K. The origin of these features is difficult to establish from macroscopic magnetization data. The small proportion of the clusters responsible for the weak 62 K transition does not result in a distinct glassy transition or a FM-like transition. This behavior of $Gd_{0.7}Ba_{0.3}MnO_3$ is attributed to the large size mismatch between the A-site cations or large $\sigma^2$ value (0.028 Å$^2$), the mismatch being considerably smaller in $Nd_{0.7}Ba_{0.3}MnO_3$ [12, 13]. Such size mismatch favours chemical/electronic inhomogeneities. To our knowledge, this is a unique case of a perovskite manganate showing a size disorder-induced spin-glass behavior, occurring in spite of the relatively large A-site cations radius ($\langle r_A \rangle$ = 1.216 Å). This behavior is comparable to the one observed in $Nd_{0.7}Ca_{0.3}CoO_3$ [20]. It appears that the so-called FM insulating state or non-magnetic insulating state often reported in the rare earth manganates of the type $Ln_{1-x}A_xMnO_3$ arises from the glassy behavior of the magnetic clusters in these materials, generally associated with electronic phase separation.

## Acknowledgement

Financial support for this work from the Swedish agencies SIDA/SAREC and VR through the Asian–Swedish research links programme is acknowledged. The authors thank BRNS (DAE), India for support of this research. AKK wants to thank University Grants Commission, India for a fellowship award and Motin for his help during samples preparation.

## References:

(1) (a) Rao C N R and Raveau B (Eds) 1998 *Colossal Magnetoresistance, Charge Ordering and Related Properties of manganese Oxides* (World Scientific: Singapore) (b) Tokura Y (Ed) 1999 *Colossal Magnetoresistance Oxides* (London: Gorden and Breach)

(2) Dagotto E (Ed) 2003 *Nanoscale Phase Separation and Colossal Magnetoresistance*; (Berlin: Springer)

(3) Uehara M, Mori S, Chen C H and Cheong S W 1999 *Nature* **399** 562

(4) Rao C N R 2000 *J. Phys. Chem. B* **104** 5877

(5) (a) Rao C N R, Kundu A K, Seikh M M and Sudheendra L 2004 *Dalton Trans.***19** 3003 (b) Rao C N R and Vanitha P V 2002 *Curr. Opin. Solid State Mater. Sci.* **6** 97

(6) Sudheendra L and Rao C N R 2003 *J. Phys.: Condens. Matter* **15** 3029

(7) Rodriguez-Martinez L M and Attfield J P 2000 *Phys. Rev. B* **63** 024424

(8) Deac I G, Diaz S V, Kim B G, Cheong S W and Schiffer P 2002 *Phys. Rev. B* **65** 174426

(9) Freitas R S, Ghivelder L, Damay F, Dias F and Cohen L F 2001 *Phys. Rev. B* **64** 144404

(10) Nam D N H, Mathieu R, Nordblad P, Khiem N V and Phuc N X 2000 *Phys. Rev. B* **62** 1027

(11) Lopez J, Lisboa-Filho P N, Passos W A C, Oritz W A, Araujo-Moreira F M, de Lima O F, Schaniel D and Ghosh K 2001 *Phys. Rev. B* **63** 224422

(12) Kundu A K, Seikh M M, Ramesha K and Rao C N R 2005 *J. Phys.: Condens. Matter* **17** 4171 and the references therein

(13) Maignan A, Martin C, Hervieu M, Raveau B and Hejtmanek J 1998 *Solid State Commun.* **107** 363

(14) Magnusson J, Djurberg C, Granberg P and Nordblad P 1997 *Rev. Sci. Instrum.* **68** 3761

(15) (a) Binder K and Young A P 1986 *Rev. Mod. Phys* **58** 801 (b) Mydosh J A 1993 *In "Spin Glasses: An Experimental Introduction"* Taylor and Francis: London

(16) Kundu A K, Nordblad P and Rao C N R 2005 *Phys. Rev. B* **72** 144423

(17) Mathieu R, Nordblad P, Nam D N H, Phuc N X and Khiem N V 2001 *Phys. Rev. B* **63** 174405

(18) Fisher D S and Huse D A 1988 *Phys. Rev. B* **38** 373

(19) (a) Itoh M, Natori I, Kubota S and Matoya K 1994 *J. Phys. Soc. Jpn.* **63** 1486 (b) Nam D N H, Jonason K, Nordblad P, Khiem N V and Phuc N X 1999 *Phys. Rev. B* **59** 4189

(20) Kundu A K, Nordblad P and Rao C N R 2006 *J. Solid State Chem.* ***179*** 923

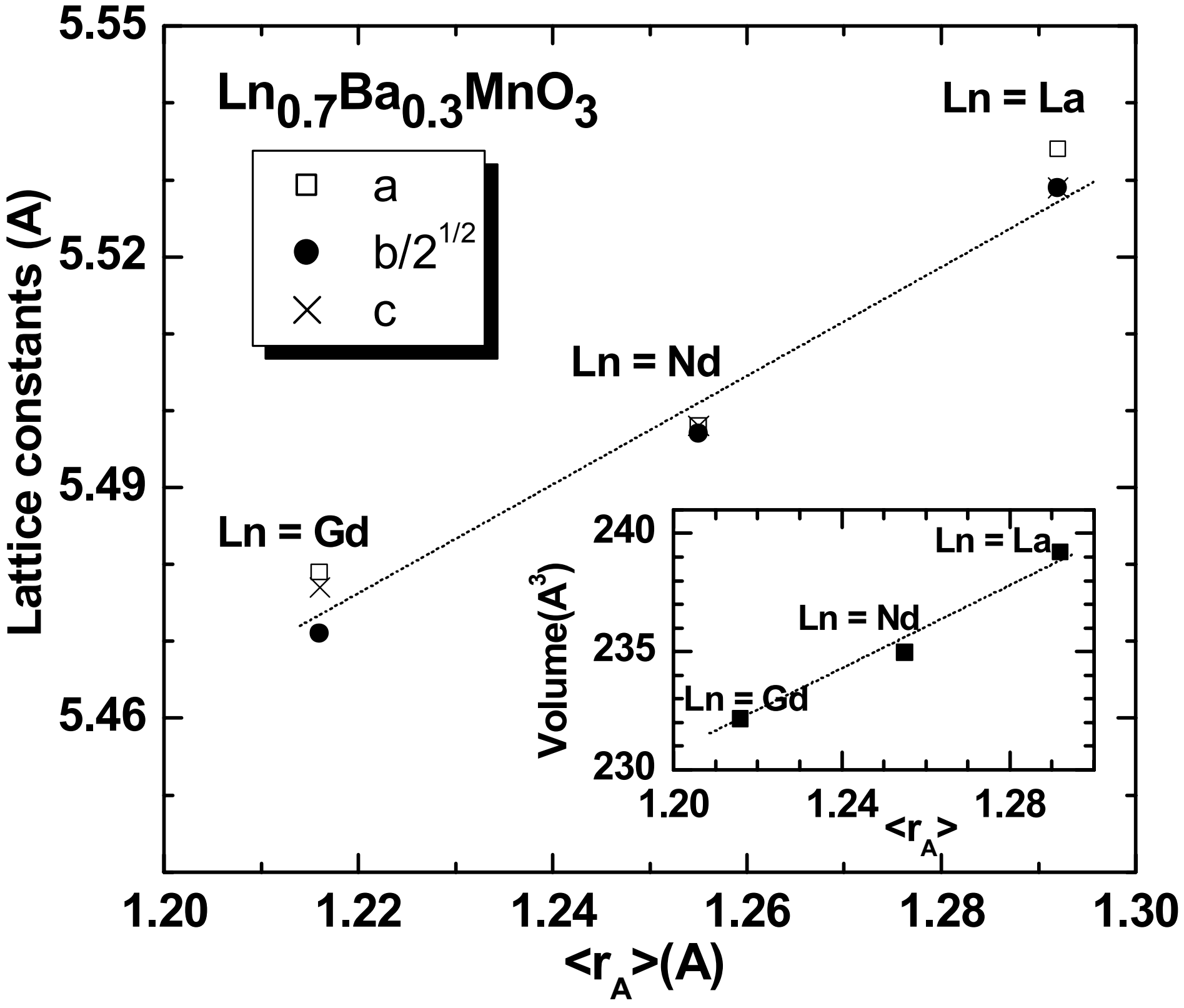

**Figure 1.** Variation of lattice parameters and cell volume (inset figure) with $<r_A>$ of $Ln_{0.7}Ba_{0.3}MnO_3$ with Ln = La, Nd and Gd.

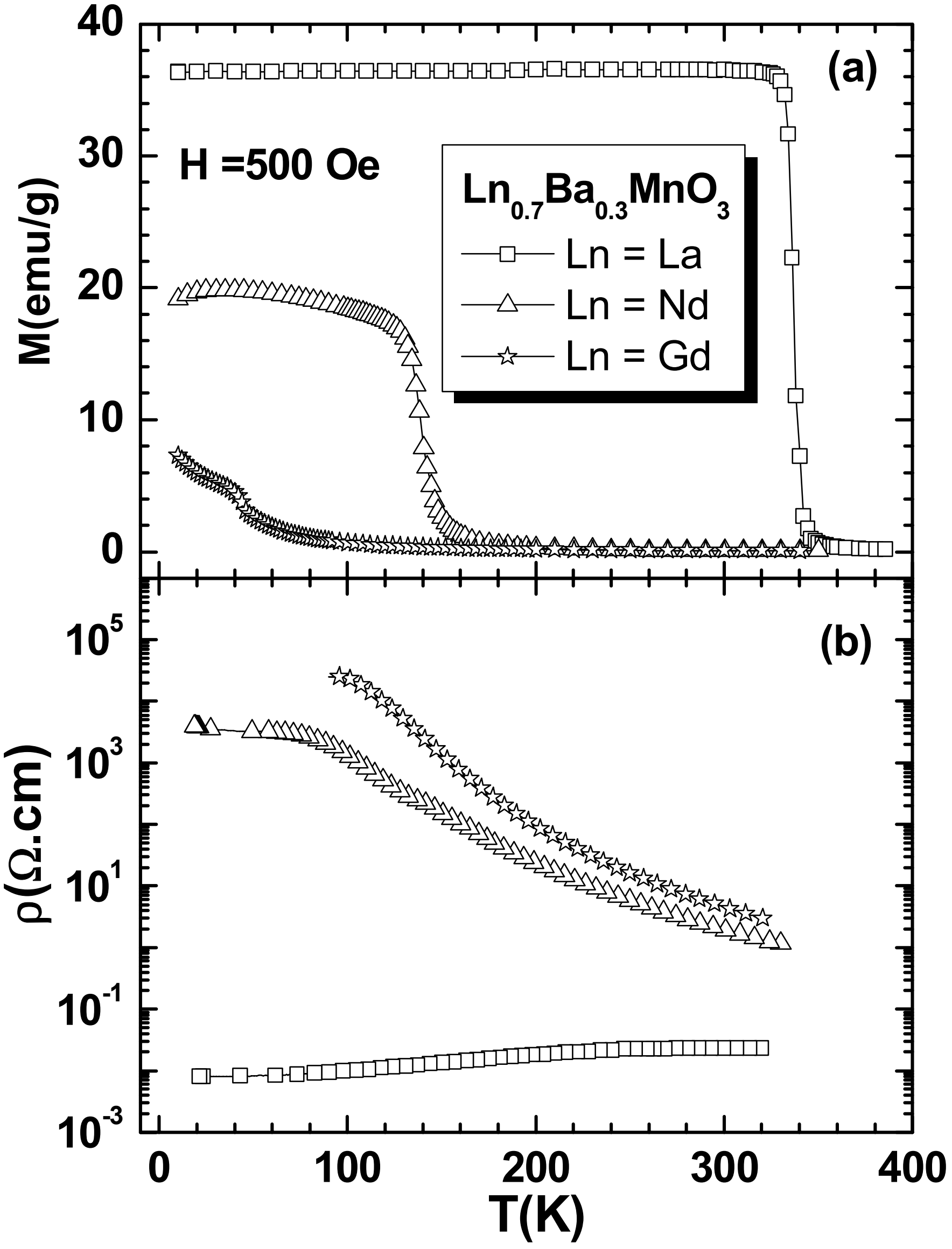

**Figure 2.** Temperature dependence of (a) the FC magnetization, M, (at H = 500 Oe) and (b) the electrical resistivity, ρ, of $Ln_{0.7}Ba_{0.3}MnO_3$ with Ln = La, Nd and Gd. Note that $Nd_{0.7}Ba_{0.3}MnO_3$ is insulating at 150 K where there is weak magnetic transition.

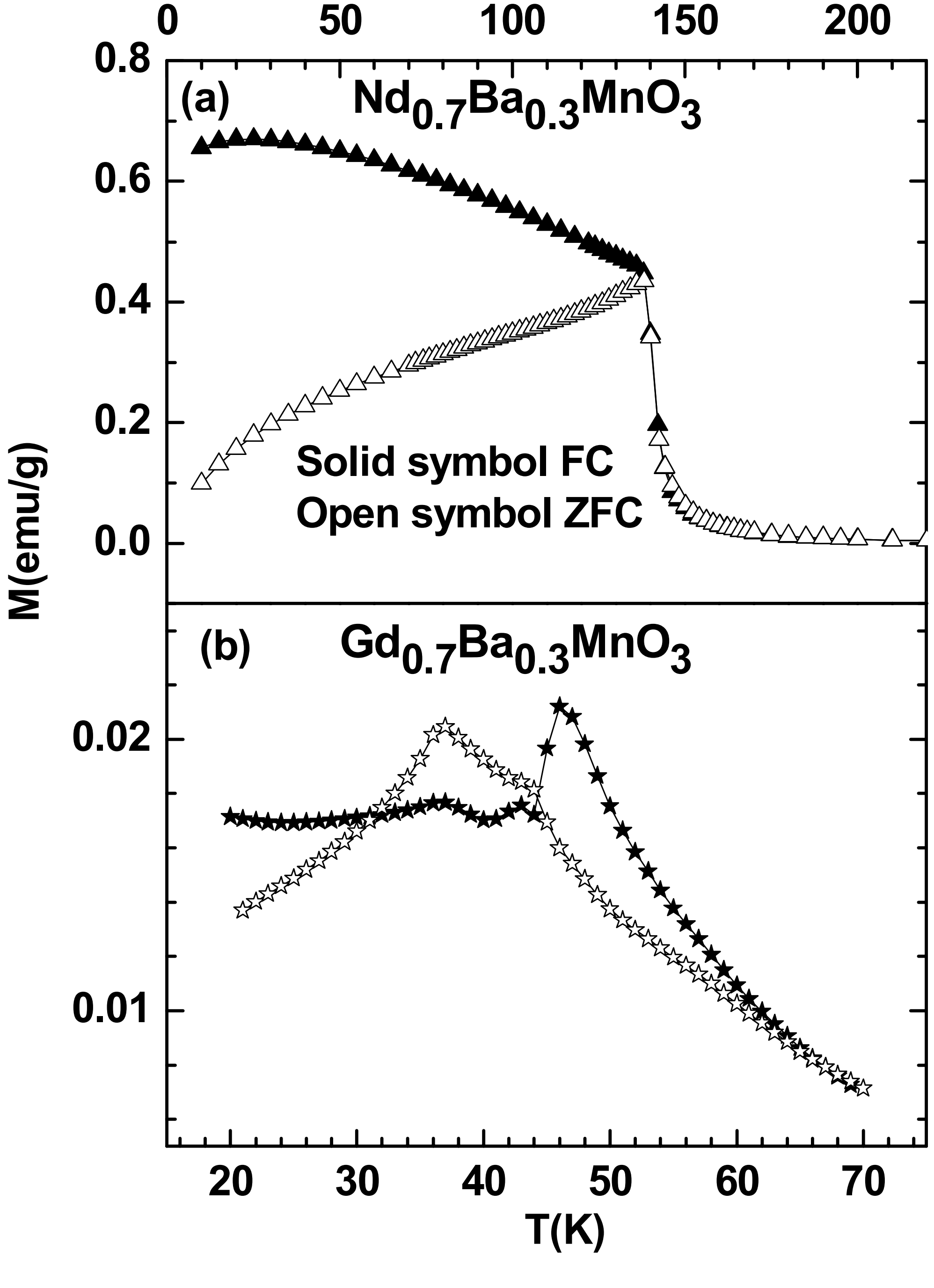

**Figure 3.** Temperature dependence of the ZFC (open symbols) and FC (solid symbols) magnetization, M, of (a) $Nd_{0.7}Ba_{0.3}MnO_3$ (at H = 10 Oe) and (b) $Gd_{0.7}Ba_{0.3}MnO_3$ (at H = 3 Oe). The features of the M-T curves remain same when the magnetic field is in the 1-10 Oe range.

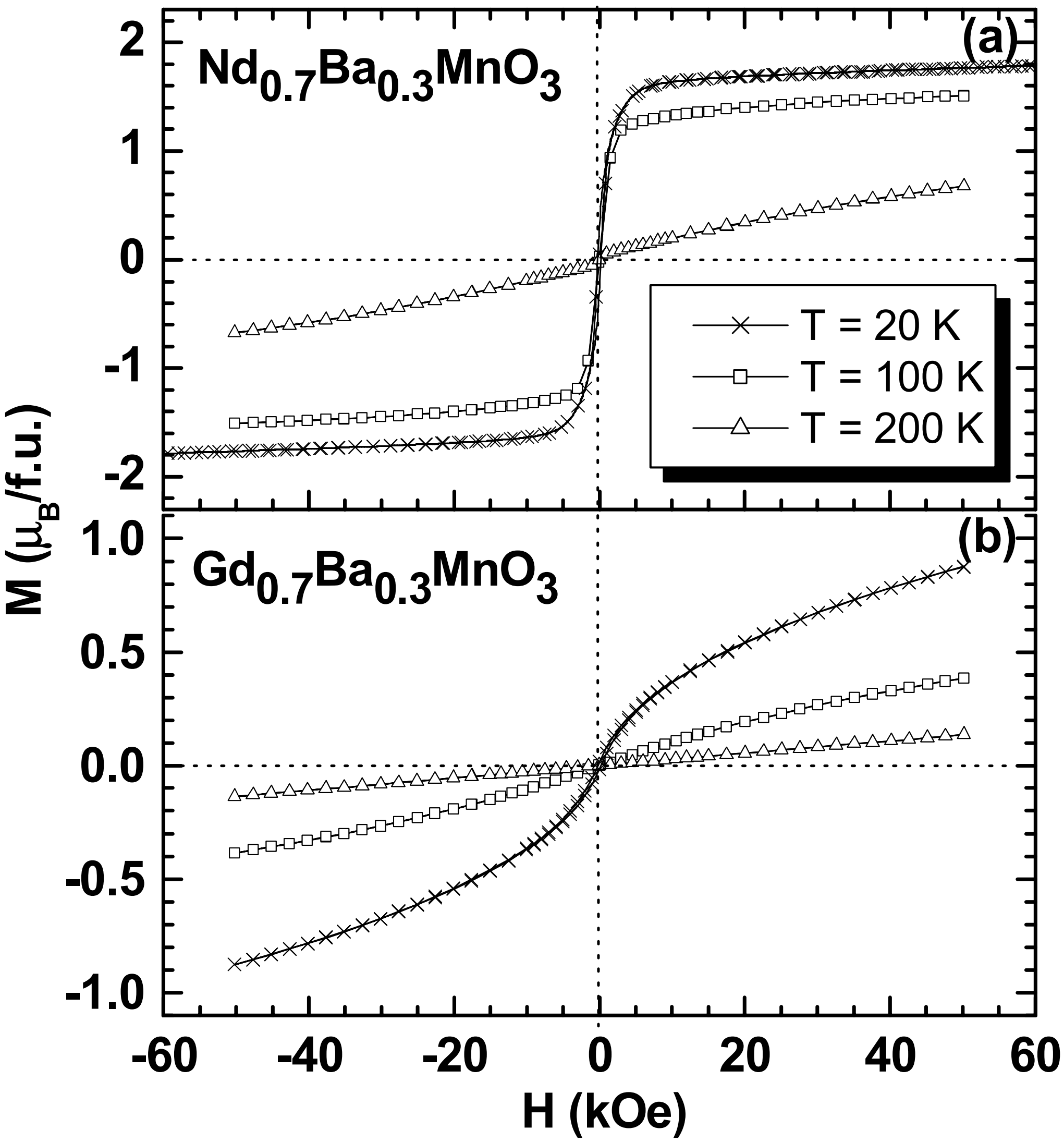

**Figure 4.** Typical hysteresis curves for (a) $Nd_{0.7}Ba_{0.3}MnO_3$ (b) $Gd_{0.7}Ba_{0.3}MnO_3$ at different temperatures.

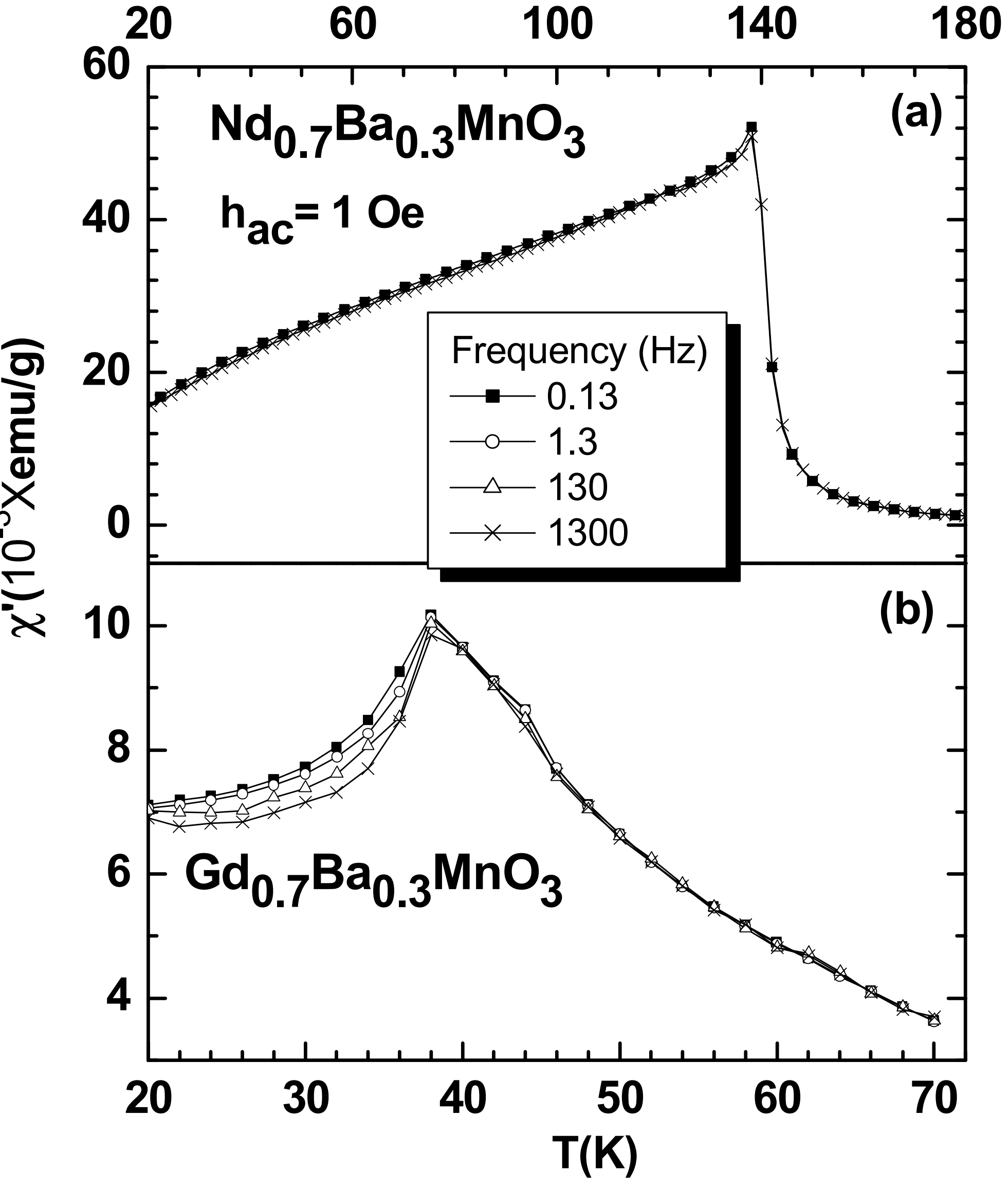

**Figure 5.** The temperature dependence of the in-phase ac-susceptibility for (a) $Nd_{0.7}Ba_{0.3}MnO_3$ and (b) $Gd_{0.7}Ba_{0.3}MnO_3$ at different frequencies.

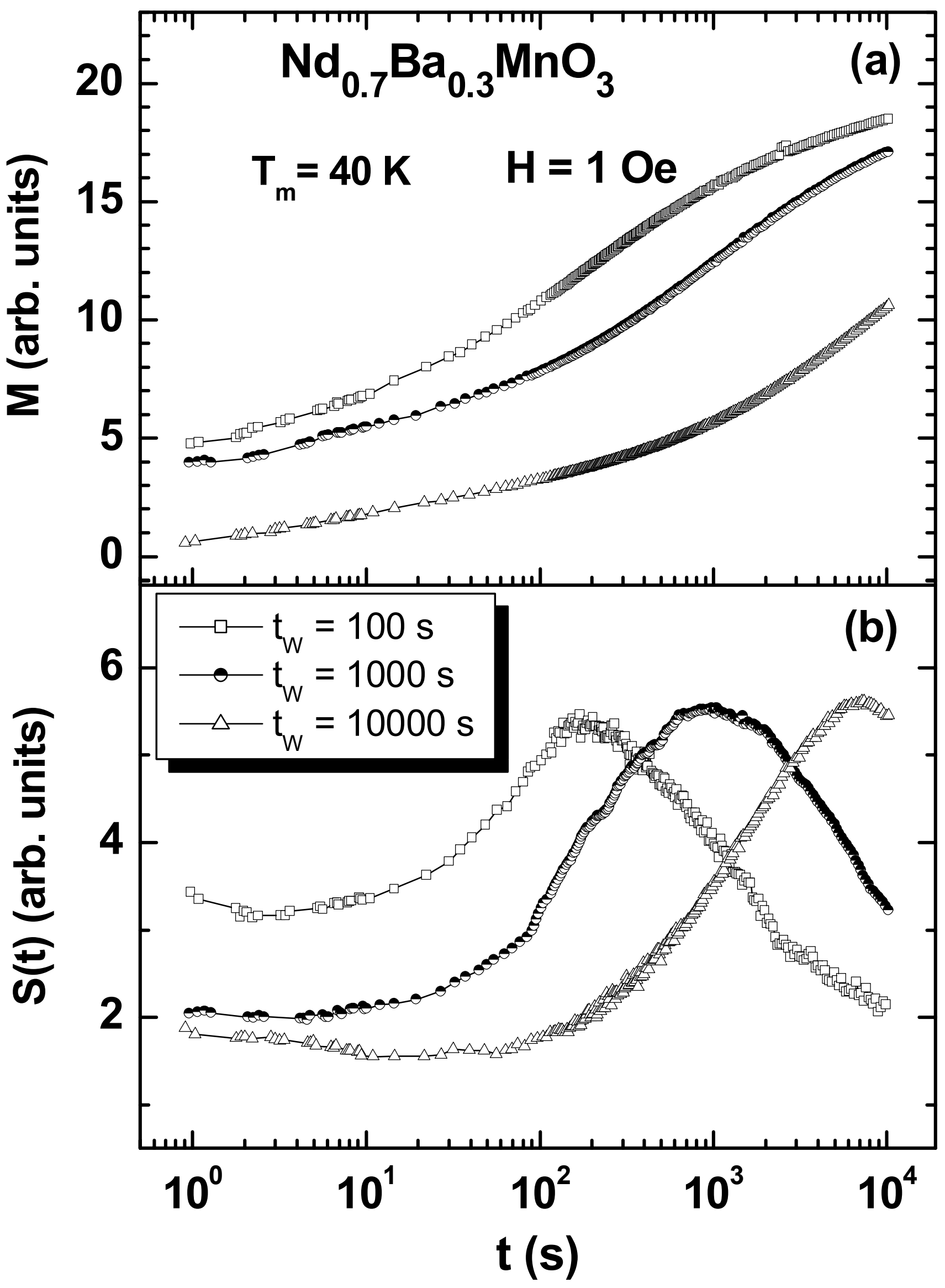

**Figure 6.** ZFC-relaxation measurements on $Nd_{0.7}Ba_{0.3}MnO_3$ at $T_m$ = 40 K for different waiting times, $t_w$ = 100, 1000 and 10000s (H = 1 Oe).

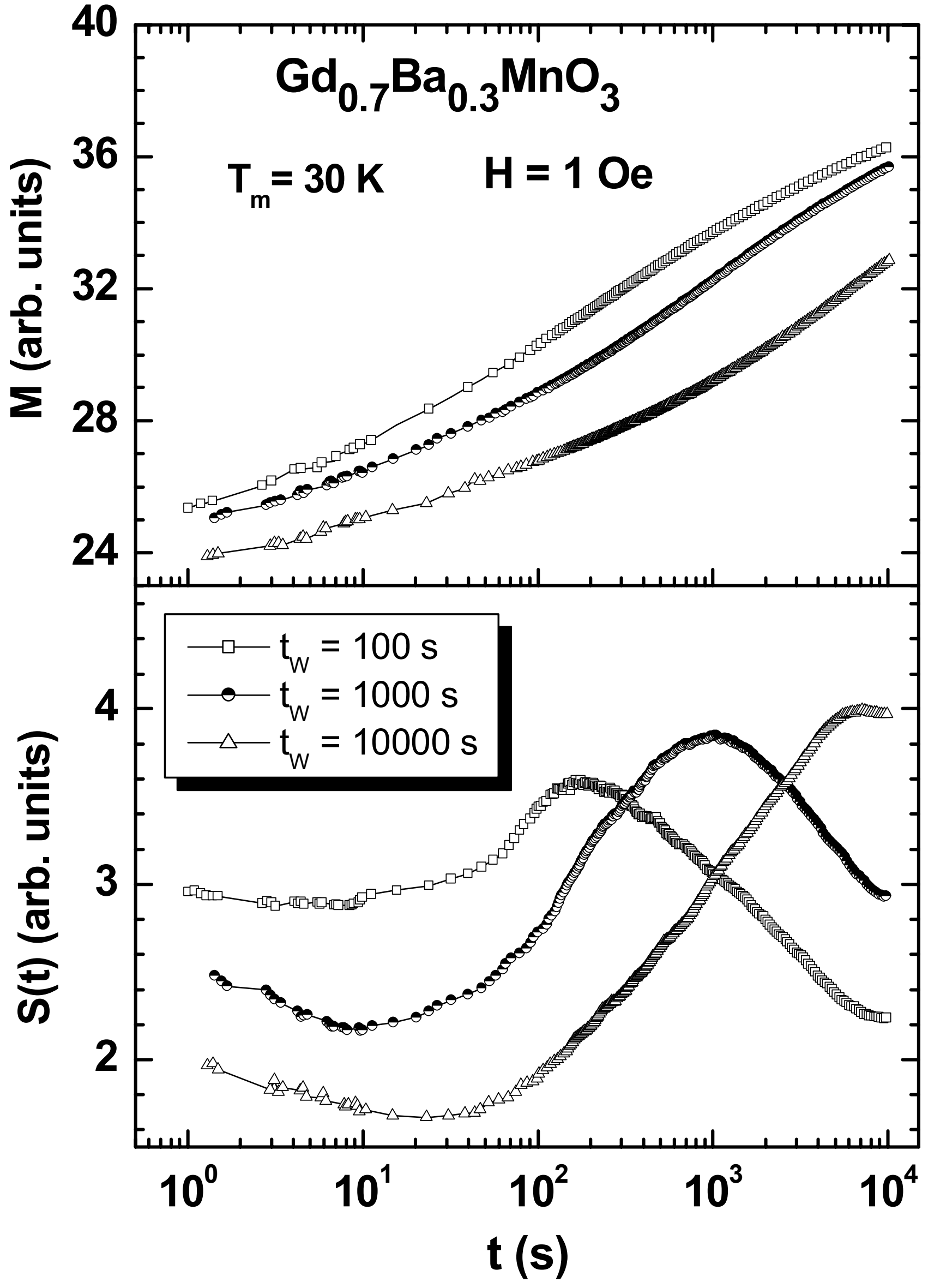

**Figure 7.** ZFC-relaxation measurements on $Gd_{0.7}Ba_{0.3}MnO_3$ at $T_m$ = 30 K for different waiting times, $t_w$ = 100, 1000 and 10000s (H = 1 Oe).

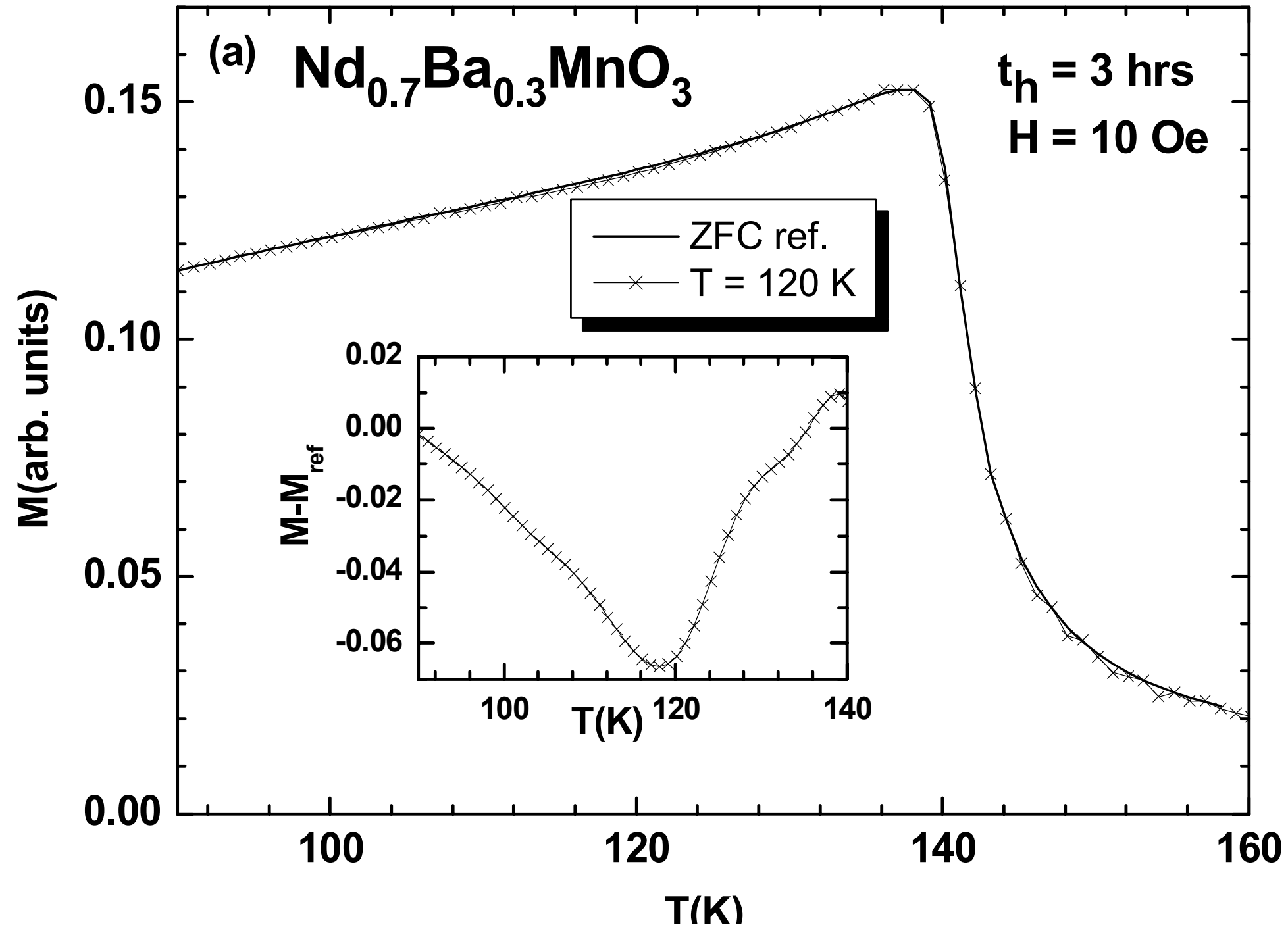

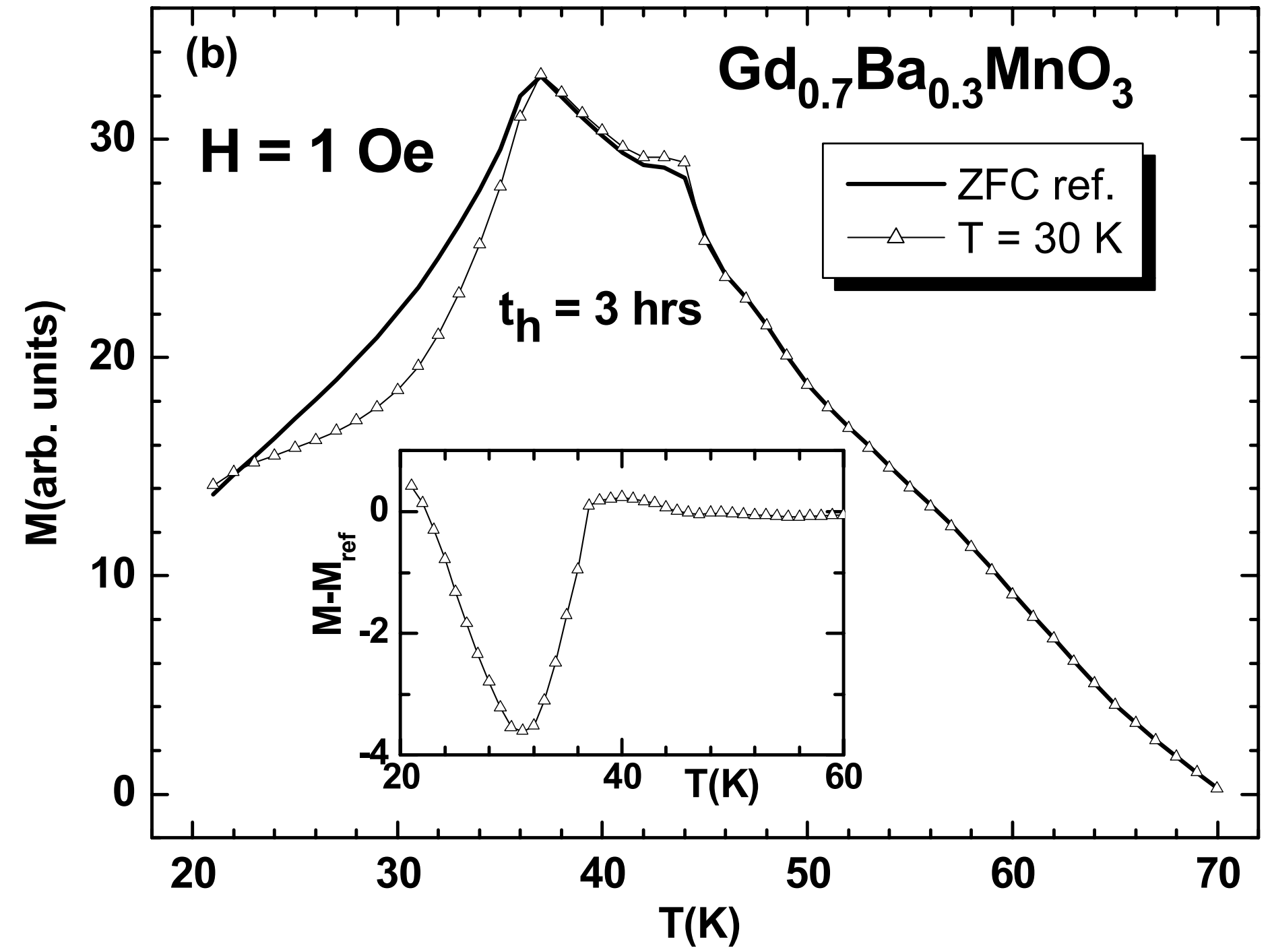

**Figure 8.** ZFC magnetization memory experiment on (a) $Nd_{0.7}Ba_{0.3}MnO_3$; the temperature dependence of ZFC magnetization, M, (reference curve) and on imprinting memories at temperature stops (120 K) during cooling for 3 hours (H = 10 Oe) and the inset shows difference ($M_{mem}$-$M_{ref}$) plot. (b) $Gd_{0.7}Ba_{0.3}MnO_3$; the temperature dependence of ZFC magnetization, M, (reference curve) and on imprinting memory at 30 K during cooling for 3 hours (H = 1 Oe). The inset shows the difference ($M_{mem}$-$M_{ref}$) plot.